\documentclass[conference, 12pt, onecolumn,draftclsnofoot]{ieeeconf}

\IEEEoverridecommandlockouts
\usepackage[usenames]{color}
\usepackage{enumerate}
\usepackage{url}
\usepackage{amsfonts,mathrsfs}
\usepackage{amssymb,amsmath}
\usepackage{verbatim}
\usepackage{acronym}
\usepackage{cite}
\usepackage{mathtools}
\usepackage{graphicx}

\usepackage{caption}

\usepackage{subfigure} 
\usepackage{epstopdf} 
\usepackage{epsfig}
\usepackage{algpseudocode,algorithm}
\usepackage[T1]{fontenc}

\usepackage{geometry}
\usepackage{dsfont}

\def\fskip#1{}

\renewcommand{\theenumi}{\arabic{enumi}}

\def\1{{\bf 1}}

\newcommand{\remove}[1]{}

\begin{document}
\title{A Novel Approach to the Behavioral Aspects of Cybersecurity}

\author{\authorblockN{Sarah Sharifi}
Department of Applied Cognition and Neuroscience \\
University of Texas at Dallas \\
sarah.sharifi@utdallas.edu
}

\maketitle
\thispagestyle{empty}
\pagestyle{plain}


\section{A Novel Approach to the Behavioral Aspects of Cybersecurity}
\label{introduction}

The Internet and cyberspace are inseparable aspects of everyone's life.
Cyberspace is a concept that describes widespread, interconnected, and online digital technology.
Cyberspace refers to the online world that is separate from everyday reality.
Since the internet and the virtual world are very recent advances in human lives, there are many unknown and unpredictable aspects to it that sometimes can be catastrophic to users in financial aspects \cite{bouveret2018cyber, walters2015cyber, popescu2018risks, tariq2018impact, quader2021insights, ghassami2017covert}, high-tech industry \cite{beluli2019smart, mcmahon2016new, amir2018firms, yekkehkhany2021adversarial, yekkehkhany2020hitting, danyk2017hybrid, gengler1999cyber}, and even life-threatening aspects in healthcare \cite{jahangiri2022carotid, muthuppalaniappan2021healthcare, alzubi2021cyber, coventry2018cybersecurity, sethuraman2020cyber, nwosu2021blockchain, hosseini2017mobile}.
Cybersecurity failures are usually caused by human errors or their lack of knowledge.
According to the International Business Machines Corporation (IBM) X-Force Threat Intelligence Index in 2020, around $8.5$ billion in records were compromised in 2019 due to failures of insiders, which is an increase of more than $200$ percent compared to the number of records that were compromised in 2018.
In another survey performed by the Ernst \& Young Global Information Security during 2018-2019, it is reported that $34\%$ percent of the organizations stated that the employees who are inattentive or do not have the necessary knowledge are the principal vulnerabilities of cybersecurity, and $22\%$ of the organizations indicated that phishing is the main threat to them~\cite{hong2021understanding}.
As stated earlier, it is noteworthy to mention that inattentive users are one of the reasons for data breaches and cyberattacks. 
In fact, the National Cyber Security Centre (NCSC) in the United Kingdom observed that $23.2$ million users who were victims of cybersecurity attacks used a carelessly selected password, which is ``123456'', as their account password.
On the other hand, the Annual Cybersecurity Report published by Cisco in 2018 announced that phishing and spear phishing emails are the root causes of a good number of cybersecurity attacks in recent years.
Given the examples above, enhancing the cybersecurity behaviors of both personal users and organizations can protect vulnerable users from cyber threats.
In fact, both human factors and technological aspects of cybersecurity should be addressed in organizations for a safer environment.

There are multiple environmental influences studied as factors that contribute to cybersecurity behaviors that emerge from the interaction of an individual operating in a cyber environment, which are listed below:
\begin{enumerate}
    \item \textbf{Organizational Culture
for Information Security Management} \cite{chang2007exploring}:
Chang and Lin mentioned that the information security technologies utilized in organizations are insufficient for information security management. Hence, although such strategies are important and necessary to be implemented whether at the intra-organizational level or across inter-organizational partners, organizations should also adopt a combination of information security and organization culture aspects.
In other words, organizations should both monitor the ``outside'' patterns and threats and the ``inside'' human nature such as relationships and activities among employees that are mostly hidden and unconscious.
The organizational culture for information security can be categorized as follows:
\begin{itemize}
    \item \textbf{Cooperativeness:} The cooperativeness trait refers to the internal culture of cooperation, trust, team work, empowerment, and information sharing inside the organizations.
    Such behavioral traits create a friendly environment among peers and colleagues and prepares a platform for all members to share information and trust each other such as an extended family.
    Internal cyberattacks in such a friendly environment is less likely.
    \item \textbf{Innovation:} Although this trait may not directly be connected to information security, an organization that promotes creativity, entrepreneurship, and adaptability will benefit from creativity of employees in all aspects including the cybersecurity domain to protect against cyberattacks.
    \item \textbf{Consistency:} The consistency trait emphasizes on order, rules and regulations, uniformity, and efficiency.
    A consistent organization is usually a formalized and regularized company in which cyberattacks and information breaches are less likely.
    \item \textbf{Effectiveness:} The effectiveness trait has an emphasis on production, goal achievement, and benefit-oriented measures.
    A company that empowers such traits in its employees would have a lower risk in cybersecurity attacks and information leak.
\end{itemize}
    The four aforementioned constructs of an organization are initiated using $26$ items that are adapted from instruments measuring the culture of a company \cite{cameron1985cultural, denison1991organizational, denison2004corporate, boggs2004tqm}.
    \item \textbf{Policies, Participation in the Security Education, Training, and Awareness Program}~\cite{han2017integrative}:
    Han et al. mention that most organizations utilize security technologies to induce the employees to comply with Information Security Policies (ISP). However, the state-of-the-art literature on ISP compliance found out that in addition to technology advances in information security, behavioral and social approaches should also be employed to avoid information breaches.
    The research results by Han et al. suggests that psychological contract fulfillment help mitigating the adverse effects of costs on the information security compliance intention in supervisor teams.
    Furthermore, employees which are aware of policies by participating in the security education, training, and awareness programs, are anticipated to act in accordance with the information security policies since they are aware of the avails of complying with the information security policies.
    This study considers a bilateral perspective that compares supervisor and supervisee groups.
    The reason for separating the supervisor groups from supervisee groups in this study is that these two groups usually have different employment qualities, they are usually in different age groups, they usually differ in their social status and work experiences. As a result, the effects of psychological contracts on their behavior in the organizations may differ.
    The results in the study by Han et al. demonstrate that the impact of psychological contract fulfillment on the information security policies compliance is more considerable for managers than for their supervisees.
    Although the work by Han et al. focuses on the psychological contract effects on information security policy compliance for supervisor groups only, there is a lot of opportunities in studying other personalized features of employees, such as job position or tenure, age, work experience, and social status to elaborate other factors that influence the information security policy compliance. In a related cross-culture study by Hovav and D'Arcy \cite{hovav2012applying}, they showed that social status, age, and gender influence the intention to misuse information security among Korean users.
    In particular, more high-ranking members of organizations who have access to more sensitive and crucial information are more plausible to breach the information security policies. Such misuses by senior members of organizations can lead to catastrophic consequences for the company.
    As a result, it is more important to enforce the perception of psychological contract fulfillment in the supervisors groups rather than the supervisee groups to reinforce the information security policy compliance intention.
    \item \textbf{Organizational Structure} \cite{hong2022motivating}: \newline
    Among factors that instigates information security breaches are behavioral and psychological characteristics.
    There are numerous studies in the literature that address the influencing factors of information security policy compliance behavior in companies. However, in the study by Hong and Furnell, they consider the influence of organizational structures.
    The authors integrate the theory of planned behavior and the perceived organizational formalization to study the procedure used to form the  information security policy compliance behavioral intentions.
    Authors use the data from a survey of organization employees, in which $261$ people take part, and the results of the data analysis utilizing the structural equation modeling is as follows.
    The empirical results suggests that the behavioral habits and cognitive processes theorized by the theory of planned behavior are notably affected by the perceived organizational formalization.
    This research study proposes that in an effort to enhance the employee information security policy compliance intentions and behavioral habits, organizations are recommended to plan a formalized set of procedures, rules, communications, and policies in general.
    Additionally, this study has the following results.
    i) Subjective norms can positively influence attitude, perceived behavioral control, and deterrence certainty, and attitude and perceived behavioral control showed strong positive influences on behavioral intention.
    ii) The decision-making process for information security policy compliance has been demonstrated to be significantly influenced by organizational formalization. Organizational formalization, in particular, had a favorable impact on attitudes toward information security behavior, perceived behavioral control, and subjective norms.
    iii) Deterrent certainty and behavioral habit can both benefit from organizational formalization.
    iv) Perceived behavioral control can encourage behavior, and deterrence certainty can encourage information security policy compliance.
    v) Although subjective norms do not have a direct impact on behavioral intention to comply with information security policies, they do influence it through the mediating effects of attitudes, deterrent certainty, and perceived behavioral control.
    \item \textbf{Managerial Participation, and Leadership} \cite{guhr2019impact, hu2012managing}:
    An important aspect of behavioral cybersecurity in organizations is the influence of management leadership on the information security behavior of employees.
    However, the importance of leadership role in the context of cybersecurity has not been explored extensively.
    This gap in the literature regarding the influence of leadership on the information security behavior of employees is addressed in the work by Guhr et al. \cite{guhr2019impact}.
    The research by Guhr et al. utilizes an interactional psychology approach to link the elements of the full-range leadership model to employees' security compliance and participation intentions. Guhr et al. evaluate a multitheoretical model on a proprietary data set that includes $322$ people from more than 14 branches across the globe. This research adds to the body of knowledge in the fields of information security and cybersecuriy by investigating the way that different leadership approaches improve the intended information security behavior of employees.
    The empirical results by Guhr et al. highlight the significance of transformational management as it can have an impact on the behavior of employee on extra-role and in-role levels regarding information security.
    In a related work by Hu et al. \cite{hu2012managing}, an individual behavioral model is developed that combines the theory of planned behavior with the role of management and the culture of the company in order to realize the effects of management on the employees' security compliance behavior.
    Hu et al. find out that the attitudes of Employees toward the perceived and subjective norms of behavioral control over compliance with information security regulations are notably influenced by top management participation in information security activities. They also discover that top management involvement has a remarkable impact on corporate culture, which in turn has an impact on the attitudes of employees about the perceived behavioral control over information security regulations. Additionally, Hu et al. discover that the impacts of management involvement and organizational culture on employee behavioral intentions are conciliated by employee cognitive assumptions regarding information security policy compliance. In addition to the deterrence-oriented treatments offered in the literature, their findings enhance the information security state-of-the-art research by showing the way that management can play a proactive role in molding the compliance behavior of employees.
\end{enumerate}

It has been demonstrated that organizational support for employees inline with information security, as a crucial environmental factor, can play an important role in enhancing productive performance of employees in regards with cyberattacks and information breaches \cite{eisenberger1986perceived}.
In a related research article, Warkentin et al. indicated that employees who have necessary access to situational support such as help from managers and colleagues, practicing behaviors, and interpersonal help, tend to be conductive to their self-efficacy regarding their information security behaviors.
An extensive study on the behavioral factors in cybersecurity is undertaken in the work by Hong and Furnell \cite{hong2021understanding} to support the hypotheses that, a) ``Behavioral comprehensiveness will have a positive impact on habits'', b) ``Self-efficacy will have a positive impact on habits'', c) ``Response efficacy will have a positive impact on habits'', d) ``Self-efficacy will have a positive impact on behavioral
comprehensiveness'', e) Response efficacy will have a positive impact on behavioral
comprehensiveness, f) ``Situational support will have a positive impact on self-efficacy'', and g) ``Situational support will have a positive impact on response efficacy''.
Hong and Furnell use cross-sectional survey to evaluate their research model.
According to the 41$^{st}$ Statistical Report on Internet Development in China, $25.4\%$ of the $772$ million internet users are students, which is the highest proportion of internet users.
Since students use internet for numerous purposes such as attending online classes and workshops, doing online shopping, interacting with social media, and sending emails and receiving emails, they are very vulnerable to cybersecurity attacks.
As a result, college students are used as the participants in Hong and Furnell's study.
The total number of $432$ students took part in the survey, out of which $393$ of the questionnaire were included for the statistical analysis.
The demography of the respondents are as follows. $200$ males $(51\%)$ and $193$ female participants $(49\%)$ responded to the questionnares, where $39$ were freshmen
(first-year students)
$(10\%)$, $108$ were sophomores (second-year students, $27\%$) , $169$ were juniors
(third-year students, $43\%$), and $77$ were seniors (fourth- and final-year students, $20\%$).
The measurements that are used in this study are explained below.
\begin{enumerate}
    \item \textbf{Situational support (SS)}:
    The SS factor is measured by adopting the approach utilized by Warkentin et al. \cite{warkentin2011influence}. The Cronbach’s alpha factor of this scale is $0.906$.
    \item \textbf{Self-efficacy (SE)}:
    The SE factor is measured by adopting the approach utilized by Bulgurcu et al. \cite{bulgurcu2010information} and Hu et al. \cite{hu2012managing}. The Cronbach’s alpha factor of this scale is $0.928$.
    \item \textbf{Response efficacy (RE)}:
    The RE factor is measured by adopting the approach utilized by Johnston \& Warkentin \cite{johnston2010fear}. The Cronbach's alpha factor of this scale is $0.915$.
    \item \textbf{Habit (HA)}:
    The HA factor is measured by adopting the approach utilized by Tsai et al. \cite{tsai2016understanding}. The Cronbach’s alpha factor of this scale is $0.925$.
    \item \textbf{Behavioral comprehensiveness (BC)}:
    The BC factor is measured by adopting the approach utilized by Limayem et al. \cite{limayem2007habit}.
    \item \textbf{Control variables}:
    Contr4ol variables such as gender, college major, grade, and other scenarios can influence the information security compliance intention. Hence, such factors are taken into account in the data analysis of this study.
\end{enumerate}

\begin{figure}
\centering
\begin{minipage}{.47\textwidth}
  \centering
  \includegraphics[width=0.9\linewidth]{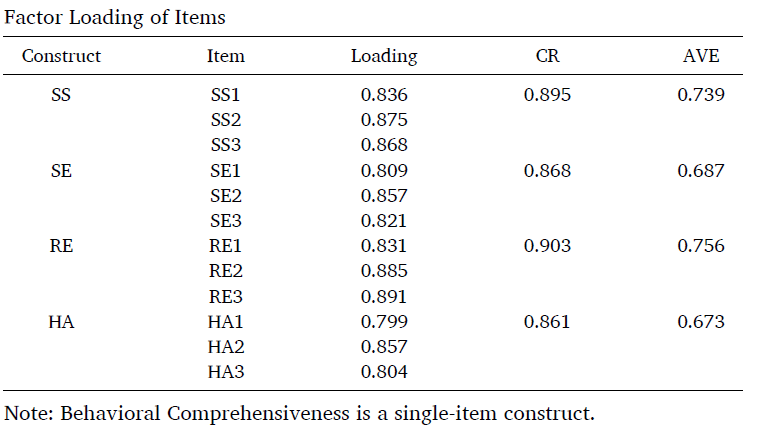}
  \captionof{figure}{Factor Loading of Items for SS, SE, RE, and HA \cite{hong2021understanding}.}
  \label{table_3}
\end{minipage}%
\begin{minipage}{.53\textwidth}
  \centering
  \includegraphics[width=1\linewidth]{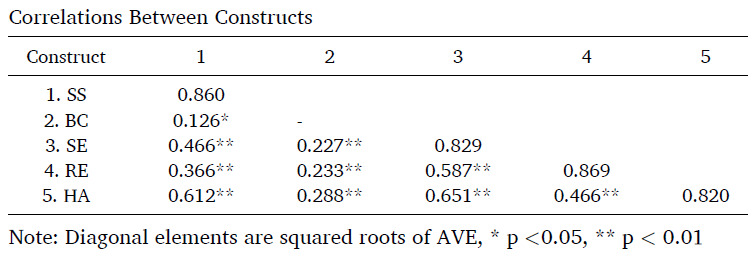}
  \captionof{figure}{The correlations between constructs \cite{hong2021understanding}.}
  \label{table_4}
\end{minipage}
\end{figure}

\begin{figure}[t]
    \centering
    \includegraphics[width=0.55\textwidth]{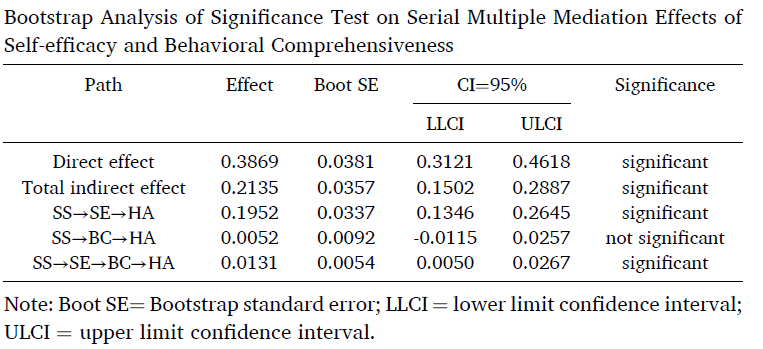}
    \caption{Bootstrap analysis of significance test on serial multiple mediation effects of self-efficacy and behavioral comprehensiveness \cite{hong2021understanding}.}
    \label{table_6}
\end{figure}

As shown in Figure \ref{table_3}, the loadings of all the items are greater than $0.5$, all the composite reliability (CR) values are greater than $0.7$, and all the
average variance extracted (AVE) values are greater than $0.5$. As a result, the assessment is considered to have a good convergent validity.
Furthermore, the correlations between constructs that are presented in Figure \ref{table_4} provide support for the hypotheses of this work.
In addition, the Bootstrap analysis of significance test on serial multiple mediation effects of self-efficacy and behavioral comprehensiveness are presented in Figure \ref{table_6}.
The data analyses support the hypotheses that both self-efficacy and response efficacy have a positive impact on behavioral comprehensiveness; situational support has a positive impact on both self-efficacy and response efficacy; and Situational support can promote cybersecurity behavioral habits through the serial multiple mediating effects of self-efficacy and behavioral comprehensiveness.


There is no doubt that breaches of data can be damaging in cyberattacks. As an example, the cyberattack on the Ukraine power grid in 2015 resulted in $225,000$ people to experience power outage \cite{maalem2020review}. It is observed that the average ransom attack jumped from $\$373$ and $\$294$ in 2014 and 2015, respectively, to $\$1077$ in 2016 \cite{maalem2020review}.
Hence, it is crucial to establish a strong curriculum in both high schools and undergraduate schools to familiarize the internet users with the consequences of cyberattacks and information theft \cite{caulkins2016cyber, frank2016early}.

\newpage

\section{Future Work}
\label{sec:conclusion}

In the context of cybersecurity and information security, one of the common themes is to approach employees of large organizations via phishing emails. In such emails, employees are incentivized with immediate rewards and tempting offers to click on links or download documents; as a result of which, an information breach occurs.
The phishing emails are usually sent to victims at the end of working hours or when they are fatigued.
As a solution to avoid information breaches as a result of employees' fatigue, two approaches are presented below:
\begin{enumerate}
    \item \textbf{Creating identifiers}:
    Most employees of large companies and organizations communicate via email internally.
    As a result, the emails that are sent from external email addresses can be labelled with an identifier so that employees pay more attention to them. As an example, the employees, including the students, professors, and staff of the University of Texas at Dallas are usually in contact with each other and send emails internally. Hence, those emails that are sent via emails that are external to the University of Texas at Dallas can be labelled for recipients.
    \item \textbf{Artificial Intelligence}:
    As mentioned earlier, employees are usually the victims of cyberattacks and phishing emails when they are fatigued.
    As a result, artificial intelligence tools such as image processing and video processing can be used to identify the level of fatigue of employees and warn them when the classification algorithms identify that the employees are fatigued.
    Machine learning methods for classification and learning can be used for fatigue detection of employees \cite{soofi2017classification, osisanwo2017supervised, kotsiantis2007supervised, kotsiantis2006machine, lecun2015deep, su2020prob2vec, yekkehkhany2019risk, king2006early, yekkehkhany2020risk, yekkehkhany2021cost, zhao2020driver}.
    Relevant methods considering the behavioral aspects of users are used in the literature \cite{mu2022research, ahmed2014robust, karuppusamy2020multimodal, lahuerta2014measuring, sun2018research, yekkehkhany2020riskthesis, yekkehkhany2020riskk, yekkehkhany2022risk, ryu2000measurement, yekkehkhany2021stochastic, gebhardt2022influence, fan2020quantitative}.
    This approach is definitely more controversial compared to the first one, since employees may not like to be monitored by image/video processing tools continuously during their work hours, even if the data is deleted and not used elsewhere. The subconscious of human beings is in a way that prefers not to be monitored, but in any case, those employees who are willing to use such a tool can benefit from the fatigue recognition aspect at the expense of being monitored.
\end{enumerate}

\bibliographystyle{IEEEtran}
\bibliography{ref}

\end{document}